\documentclass[12pt]{iopart}

\usepackage{iopams}

\usepackage{lineno,hyperref}
\modulolinenumbers[5]

\usepackage{bm}
\usepackage{color}
\usepackage{dcolumn}
\usepackage{ulem}
\usepackage{graphicx}

\def\bcero{\bn{0}}
\newcommand{\bn}[1]{\mbox{\boldmath $#1$}}

\def\bB{\bn{B}}
\def\bP{\bn{P}}
\def\bY{\bn{Y}}

\def\bW{\bn{W}}

\def\bI{\bn{I}}
\def\bA{\bn{A}}

\def\bK{\bn{K}}

\def\bH{\bn{H}}
\def\bT{\bn{T}}

\def\bF{\bn{F}}

\begin{document}

\title[]{Hybrid matrix method for stable numerical analysis of the propagation of Dirac electrons in gapless bilayer graphene superlattices}

\author{J. A. Briones-Torres$^1$, R. Pernas-Salom\'{o}n$^{2,3}$, R. P\'{e}rez-\'{A}lvarez$^3$ and I. Rodr\'{i}guez-Vargas$^1$}
\address{$^1$Unidad Acad\'{e}mica de F\'{i}sica, Universidad Aut\'{o}noma de Zacatecas, Calzada Solidaridad Esquina Con Paseo La Bufa S/N, 98060, Zacatecas, Zac., M\'{e}xico.}
\address{$^2$Instituto de F\'{\i}sica, Benem\'{e}rita Universidad Aut\'{o}noma de Puebla, 18 Sur y San Claudio, Edif. 110-A, C.P. 72570, Puebla, M\'{e}xico.}
\address{$^3$Facultad de Ciencias, Universidad Aut\'onoma del Estado de Morelos, Av. Universidad 1001, CP 62209, Cuernavaca, Morelos, M\'{e}xico.}

\ead{jabt0123@gmail.com}
\vspace{10pt}
\begin{indented}
\item[]December 2015
\end{indented}

\begin{abstract}
Gapless bilayer graphene (GBG), like monolayer graphene, is a material system with unique properties, such as anti-Klein tunneling and intrinsic Fano resonances. These properties rely on the gapless parabolic dispersion relation and the chiral nature of bilayer graphene electrons. In addition, propagating and evanescent electron states coexist inherently in this material, giving rise to these exotic properties. In this sense, bilayer graphene is unique, since in most material systems in which Fano resonance phenomena are manifested an external source that provides extended states is required. However, from a numerical standpoint, the presence of evanescent-divergent states in the eigenfunctions linear superposition representing the Dirac spinors, leads to a numerical degradation (the so called $\Omega$d problem) in the practical applications of the standard Coefficient Transfer Matrix ($\bK$) method used to study charge transport properties in Bilayer Graphen based multi-barrier systems. We present here a straightforward procedure based in the hybrid compliance-stiffness matrix method ($\bH$) that can overcome this numerical degradation. Our results show that in contrast to standard matrix method, the proposed $\bH$ method is suitable to study the transmission and transport properties of electrons in GBG superlattice since it remains numerically stable regardless the size of the superlattice and the range of values taken by the input parameters: the energy and angle of the incident electrons, the barrier height and the thickness and number of barriers. We show that the matrix determinant can be used as a test of the numerical accuracy in real calculations.
\end{abstract}

%
\vspace{2pc}

\submitto{\MSMSE}
%
\maketitle
%
%


\section{Introduction}

Since its discovery, graphene has been the subject of considerable research \cite{1,2,3,4,5,6,7,8,9,10,11,12,13,14,15,16,17}, including the propagation of electrons through multilayer systems \cite{6,7,8,9,10,11,12,13,14,15,16,17}. Akin to monolayer graphene, GBG has become a very interesting research material \cite{18,19,20,21,22,23,24}. In particular, at low energies, the properties of GBG are described by massive chiral quasiparticles that obey a quadratic dispersion relation. This opens an important door to the path of electronics because bilayer graphene is intrinsically an undoped gapless semiconductor \cite{22,23,24}, and both the carrier density and the energy bandgap can be controlled by doping or gating \cite{25,26,27}.\\

One of the most remarkable differences of gapless bilayer graphene as compared with gapless monolayer graphene is that propagating and evanescent states coexist inherently in this material, giving rise to quite interesting and exotic phenomena such as anti-Klein tunneling \cite{6} and Fano resonances  \cite{28,29,30}. The case of Fano resonances in GBG is unique, since in practically all material systems in which this phenomenon is manifested an external source that provides extended states is required \cite{31,32}. This dichotomy, the propagating and evanescent character, of Dirac electrons in GBG turns out in an asymmetric line shape of the transmission probability, when the material is subjected to an electrostatic barrier. Even, the signatures of the Fano resonances can be manifested in the transport properties of GBG, opening the possibility of testing this quite exotic phenomenon in conductance measurements. Even more, Fano resonances are appealing from the technological standpoint, since their asymmetric line shape gives the possibility of modulate the transmission properties from a negligible value to a high value in a very narrow energy or frequency range. This makes GBG an excellent candidate for electronics applications as sensors, lasing and nonlinear devices \cite{31,32}.\\

From the formal point of view the Coefficient Transfer Matrix method ($\bK$) is suitable to
study charge transport properties (e.g., transmission coefficients, conductance, etc.) in Bilayer Graphene based electrostatic single and multi-barrier systems \cite{33,34,35}. This method is so popular due to its simplicity as well as its adaptability to a wide range of wave-propagation phenomena. However, the practical application of this matrix method in GBG superlattices is hampered by a numerical degradation, the so called: the $\Omega d$ problem \cite{36,37}. The name associated with this numerical instability derives from the elastic waves studies where this instability is present at high frequencies $\Omega$ and/or
big thicknesses ($d$) of the structure layers. This is because the matrix elements contain a mixture of exponentially growing and decaying terms that lead to loss of precision during computations \cite{38,39,40,41,42}. In GBG this mixture is given by the presence of evanescent-divergent states in the eigenfunctions linear superposition representing the Dirac spinors. Due to the peculiarities presented in this material \cite{6,28,29,30} the $\Omega d$ problem is more likely in GBG superlattice studies where parameters such as the energy and angle of the incidence electrons as well as the thickness and number of barriers is increased. Likewise, in aperiodic or quasi-periodic structures this numerical degradation is natural, since the size of these structures increases as a function of the generation number \cite{41,42}.\\

Different techniques have been proposed to avoid the $\Omega d$ problem. Some of them, as the global transfer matrix involve matrices with dimensions increasing with the number of layers forming the system. Other approaches employ transfer matrices with dimensions independent of the
number of layers. Among them we can find the scattering matrix method, the impedance matrix method, the compound matrix, and the hybrid compliance-stiffness matrix or simply the hybrid matrix \cite{38,39,40,41,42}. From all these possibilities, the hybrid matrix method is an excellent option, since it is numerically stable, well-conditioned, and accurate irrespective of the thickness (large, small or even zero) of the system \cite{38,39,37}.\\

In this work we study transmission properties of electrons in GBG superlattice using the transfer matrix method. We show that the practical application of the common coefficient transfer matrix $(\bK)$ may leads to numerical instabilities (the $\Omega d$ problem) when the electron energy, the angle of incidence, the height of the barriers and/or the size of the structures grows. To overcome this numerical problem we resorted to the hybrid matrix method using the straightforward methodology addressed in \cite{37}. To this end we transform the effective Hamiltonian describing the motion of charge carriers in bilayer graphene into an ordinary second order differential system that take the Sturm-Liouville form \cite{36}. Our results show that the hybrid matrix method gives reliable outputs for transmission properties of electrons in GBG superlattice, irrespective of the system size.

\section{Methodology}

\subsection{Electrostatic field effect in gapless bilayer graphene}

\begin{figure}
\centering
\includegraphics[scale=0.2]{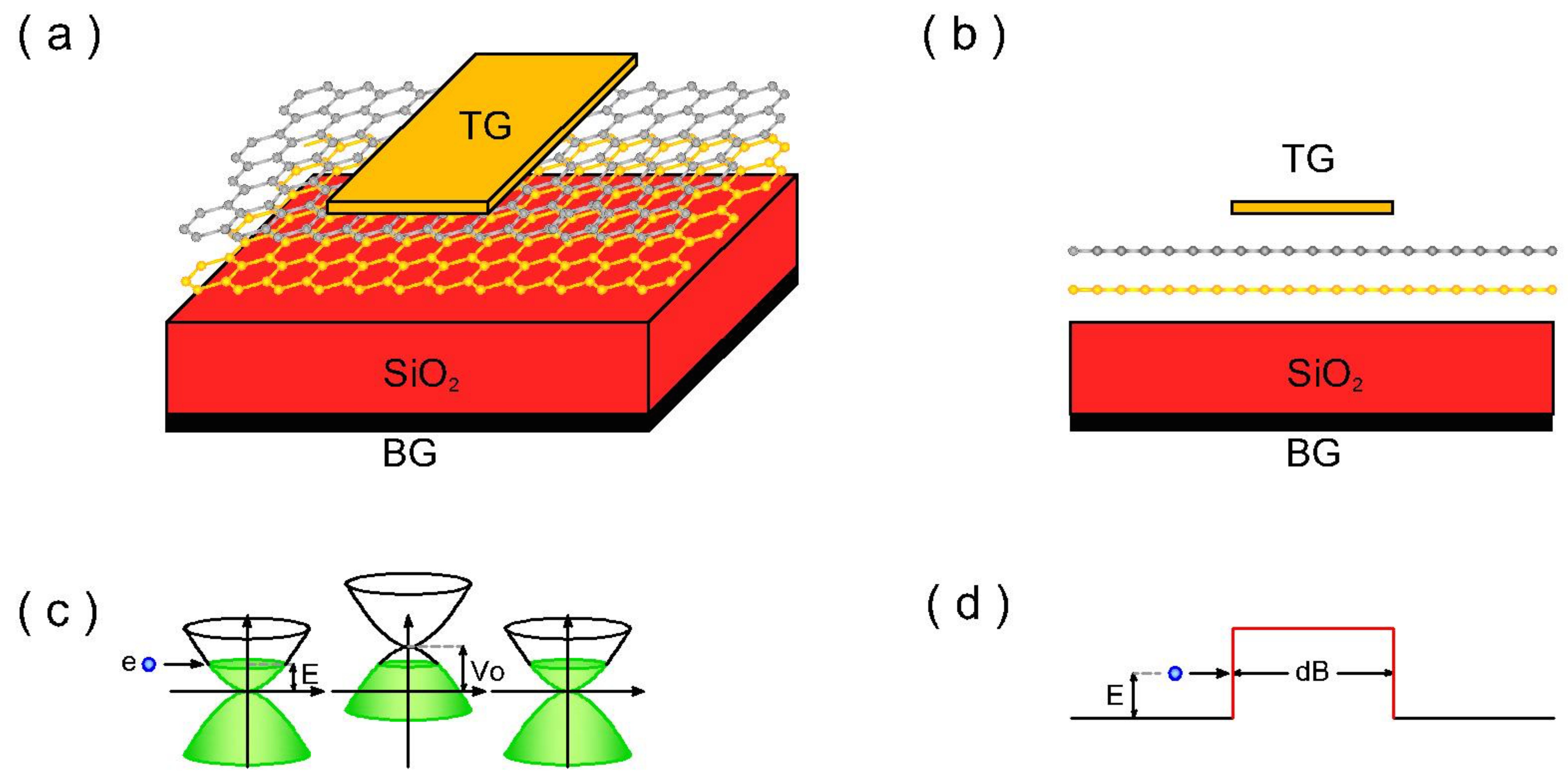}
\caption{\label{Fig1} Schematic representation (a) of the possible experimental design, (b) side view of (a), (c) dispersion-relation distribution, (d) energetic representation for a gapless bilayer graphene barrier. }
\end{figure}

\begin{figure}
\centering
\includegraphics[scale=0.25]{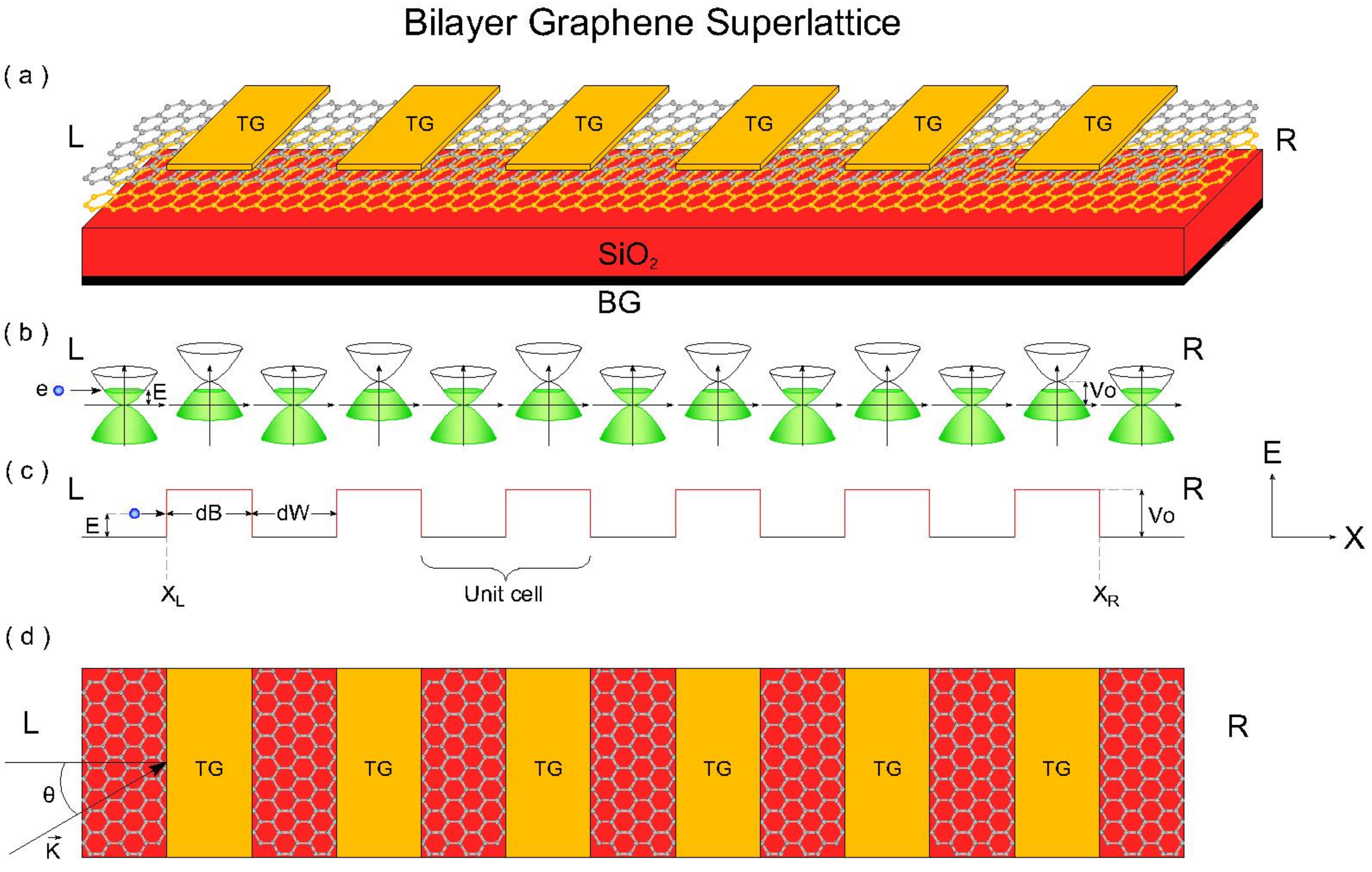}
\caption{\label{Fig2} Schematic diagram for (a) the GBG superlattice design, showing external domains L (left) and R (right), (b) dispersion-relation distribution, (c) the potential profile showing the barriers height, $V_0$ and the well/barrier width, dW/dB, and (d) top-view for the GBG superlattice design, showing the angle of incidence, $\theta$.}
\end{figure}

Our basic system is a graphene-based device, see Figure \ref{Fig1}. This consists of two graphene sheets sited on a non-breaking symmetry substrate like $\mathrm{SiO}_2$. A back gate (BG) controls the Fermi energy of Dirac electrons and a top gate (TG) suspended at a certain distance from the bilayer graphene controls the width (dB) and the height $(V_0)$ of the electrostatic barrier. Two conducting leads are attached at right and left ends of the bilayer graphene structure. The general scheme of our multi-barrier system (the GBG superlattice) in shown in Fig. \ref{Fig2}, where the one-dimensional potential profile (Fig. \ref{Fig2}c) along the x-axis is given by:

\begin{eqnarray}
    V(x) = \left\{\begin{array}{ll}
               V_0 & \mathrm{for}\;\; \mathrm{barriers} \\
               0 & \mathrm{for}\;\; \mathrm{wells }
             \end{array}\right.
 \end{eqnarray}

The charge carriers in the bilayer garphene have a parabolic energy spectrum and are described by the following Hamiltonian \cite{6,18,21}:

\begin{eqnarray}\label{Hamiltoniano}
  H = -\frac{\hbar^2}{2m}\left(
                             \begin{array}{cc}
                               0 & (q_x-iq_y)^2 \\
                               (q_x+iq_y)^2 & 0 \\
                             \end{array} \right) + V(x)\left(
                                                         \begin{array}{cc}
                                                           1 & 0 \\
                                                           0 & 1 \\
                                                         \end{array} \right),
\end{eqnarray}
where $q_x$ and $q_y$ are the quasiparticle wavevectors along the $x$ and $y$ directions respectively; $m$ is the band effective mass taken as
$0.035\; m_0$, and $m_0$ is the bare electron mass \cite{8,28,30}.\\

The corresponding eigenvalue equation:
\begin{equation}\label{Ec-Schrodinger}
    H\psi=E\psi,
\end{equation}
\noindent can be readily solved giving rise to a parabolic dispersion relation:
\begin{equation}\label{dispersion-relation}
   E-V_0= \pm \frac{\hbar^2}{2m}\mathbf{q}^2;\;\; \mathbf{q} = (q_x,q_y),
\end{equation}
\noindent and to the set of eigenfunctions:
\begin{eqnarray}\label{Eigenfunction-q}
  \psi_{\pm}^q = \left(
                             \begin{array}{c}
                               1 \\
                               v_{\pm}^q \\
                             \end{array} \right) e^{\pm iq_x x + iq_y y},
\end{eqnarray}
\noindent corresponding to the propagating wavefunctions, and
\begin{eqnarray}\label{Eigenfunction-beta}
  \psi_{\pm}^\beta =  \left(
                             \begin{array}{c}
                               1 \\
                               v_{\pm}^\beta \\
                             \end{array} \right) e^{\pm \beta_x x + iq_y y},
\end{eqnarray}
\noindent corresponding to the evanescent-divergent ones. Expressions for $q_x$ and $\beta_x$ were obtained from (\ref{dispersion-relation}):
\begin{equation}\label{}
    q_x=\sqrt{\left(\frac{2m}{\hbar^2}\right)(E-V_0)-q_y^2};
\end{equation}
\begin{equation}\label{}
    \beta_x=\sqrt{\left(\frac{2m}{\hbar^2}\right)(E-V_0)+q_y^2},
\end{equation}
\noindent with $q_y=\sin (\theta) \sqrt{2mE/\hbar^2}$. The coefficients $v_{\pm}^q$ and $v_{\pm}^\beta$ come as:
\begin{equation}\label{}
    v_{\pm}^q=-\frac{\hbar^2}{2m}\frac{\left(\pm q_x+iq_y\right)^{2}}{E-V_0},
\end{equation}
\noindent and
\begin{equation}\label{}
    v_{\pm}^\beta=-\frac{\hbar^2}{2m}\frac{\left(\pm \beta_x+q_y\right)^{2}}{E-V_0}.
\end{equation}

Here, it is important to mention that the solutions in the regions without electrostatic potentials can be straightforwardly obtained by simply setting $V_0 = 0$.\\

Finally, the wavefunction solution of Equation (\ref{Ec-Schrodinger}) for a barrier or a well domain, can be expressed by means of a linear combination of the four eigenfunctions:
\begin{equation}\label{Solucion-wavefunction}
    \psi(x,y)= A_{+} \psi_{+}^q + A_{-} \psi_{-}^q + B_{+} \psi_{+}^\beta + B_{-} \psi_{-}^\beta,
\end{equation}

\noindent where $A_{+}$, $A_{-}$, $B_{+}$, $B_{-}$ is a set of expansion coefficients.

\subsection{The Coefficient Transfer Matrix $(\bK)$}

Having obtained the eigenfunctions, the coefficient transfer matrix method can be formulated applying the proper boundary conditions at the interfaces of the structure, see for example references [40,41,42]. In our case the continuity conditions for the wavefunction (\ref{Solucion-wavefunction}) and its derivative, along the $x$ axis, are well supported. They are based on the conservation of probability current density and the transversal momentum $(q_y)$ conservation. The latter is because the $y$-component of quasi-particle momentum commutes with the Hamiltonian \cite{43}.\\

Applying the above mentioned continuity conditions at each interface of the GBG superlattice of the Figure \ref{Fig2} we obtained the coefficient transfer matrix $(\bK)$ that relates the expansion coefficients of the output wavefunction $\psi(x_R,y)$ in the semi-infinite external domain $(\mathrm{R})$ with those corresponding to the incident state $\psi(x_L,y)$ in the semi-infinite external domain $(\mathrm{L})$:
\begin{eqnarray}\label{Ec-matricial-K}
  \left( \begin{array}{l}
      A_{+}(\mathrm{L}) \\
      A_{-}(\mathrm{L}) \\
      B_{+}(\mathrm{L}) \\
      B_{-}(\mathrm{L}) \\
    \end{array} \right)  = \bK\cdot \left( \begin{array}{l}
                                          A_{+}(\mathrm{R})\;e^{i \kappa_x\, x_R} \\
                                          A_{-}(\mathrm{R})\;e^{-i \kappa_x\, x_R} \\
                                          B_{+}(\mathrm{R})\;e^{\alpha_x\, x_R} \\
                                          B_{-}(\mathrm{R})\;e^{-\alpha_x\, x_R} \\
                                          \end{array} \right);
\end{eqnarray}
\begin{eqnarray}\label{Matrix-K}
   \bK  = D_{\mathrm{L}}^{-1}\cdot \prod_{j=1}^{N} \left(D_j\cdot P_j\cdot D_j^{-1}\right)\cdot D_{\mathrm{L}},
\end{eqnarray}
\noindent where $N$ is the total number of barriers and well located between the external domains $(\mathrm{L}, \mathrm{R})$ of the GBG superlattice and $\kappa_x$ / $\alpha_x$ is the wavevector $q_x$ / $\beta_x$ evaluated for $V_0=0$. The matrices $D$ and $P$ are given by:
\begin{eqnarray}
  D = \left(  \begin{array}{cccc}
            1 & 1 & 1 & 1 \\
            v_{+}^q & v_{-}^q & v_{+}^\beta & v_{-}^\beta \\
            iq_x & -iq_x & \beta_x & -\beta_x \\
            iq_x\;v_{+}^q & -iq_x\;v_{-}^q & \beta_x\;v_{+}^\beta & -\beta_x\;v_{-}^\beta \\
          \end{array}  \right);
\end{eqnarray}
\begin{eqnarray}\label{matrix-P}
  P = \left( \begin{array}{cccc}
            e^{-iq_x\;d} & 0 & 0 & 0 \\
            0 & e^{iq_x\;d} & 0 &0 \\
            0 & 0 & e^{-\beta_x\;d} & 0 \\
            0 & 0 & 0 & e^{\beta_x\;d} \\
          \end{array} \right).
\end{eqnarray}
The subscript $\mathrm{L}$ / $j$ in the matrices $D$ and $P$ indicates the domain of the structure at which they are evaluated, ie. the subindex $\mathrm{L}$ corresponds to the semi-infinite external domain $(\mathrm{L})$, and $j=1,\ldots,N$ labels the internal domains of the GBG superlattices starting from the left of the structure. Parameter $d$ is the domain width.\\

In our problem we assume incidence from the left of the structure and we set $A_{-}(\mathrm{R})=0$ since there is no reflected wave in the right external domain $(\mathrm{R})$. In addition to this, we impose $B_{+}(\mathrm{R})=0$ and $B_{-}(\mathrm{L})=0$ to avoid nonphysical solutions as $x\rightarrow\pm\infty$. Taken into account this boundary conditions in the expression (\ref{Ec-matricial-K}) we can easily calculate the transmittance $T$ through the equation:
\begin{equation}\label{transmittance}
    T=\left|\frac{A_{+}(\mathrm{R})\;e^{i \kappa_x\, x_R}}{A_{+}(\mathrm{L})}\right|^2=\frac{|\bK_{44}|^2}{\left|\bK_{11}\bK_{44}-\bK_{14}\bK_{41}\right|^2}.
\end{equation}
As can be seen, the transmittance can be calculated using only 4 of the 16 matrix elements of the coefficient transfer matrix $(\bK)$.\\

At this point we draw attention to the value of the determinant of the matrix $\bK$ expressed in (\ref{Matrix-K}). Considering (\ref{matrix-P}) it can be checked directly from (\ref{Matrix-K}) that the determinant of the matrix $\bK$ is one. In the analysis of our results we used the value of this determinant to check the numerical accuracy of our calculations. When the numerical instability is present the determinant takes values quite different from exactly one, being in some cases several orders of magnitude bigger or smaller than 1.0.\\

Tracking the numerical stability of real calculations through the numerical value of the matrix determinant can be also applied to other type of transfer matrices. In Reference \cite{36} the determinant of the standard transfer matrix, denoted there by $\bT$, was calculated analytically using the coefficients of the Matrix Sturm-Liouville Equation (MSLE) and it was shown that its value equals unity or it is at least unimodular for a wide class of physical problems.

\subsection{The Hybrid matrix $(\bH)$ derived from the MSLE }

Taking $q_x=-i \frac{d}{dx}$ in equations (\ref{Hamiltoniano}-\ref{Ec-Schrodinger}) we can obtain an ordinary second order differential system for the unknown $\psi(x,y)$:

\begin{eqnarray}\label{solution-1}
  \psi(x,y) = \left|
              \begin{array}{c}
                \psi_1(x,y) \\
                \psi_2(x,y) \\
              \end{array}
            \right|,
\end{eqnarray}

\noindent where $\psi_1(x,y)$/ $\psi_2(x,y)$ is the two component pseudo-spin wavefunction for the graphene sublattice A / B \cite{21,28}. Considering the conservation of the transversal momentum $(q_y)$ the wavefunction (\ref{solution-1}) can be expressed as:

\begin{eqnarray}
  \psi(x,y) = \left|
              \begin{array}{c}
                \psi_1(x,y) \\
                \psi_2(x,y) \\
              \end{array}
            \right| = \left|
                        \begin{array}{c}
                          \psi_1(x) \\
                          \psi_2(x) \\
                        \end{array}
                      \right|\;e^{iq_y y},
\end{eqnarray}
\noindent then the result is an ordinary second order differential system that take the Sturm-Liouville form \cite{36}:
\begin{eqnarray}
  \label{Eqmaestra} {\mathbf{L}}(x)\cdot \bF(x)\equiv \nonumber\\
\end{eqnarray}
\begin{eqnarray}
\fl
  \frac{d}{dx} \left[ \bB(x) \cdot \frac{d\bF(x)}{dx} + \bP(x) \cdot \bF(x)\right] + \bY(x)
\cdot \frac{d\bF(x)}{dx} + \bW(x) \cdot \bF(x) &=& \bcero_{2\times 1} \;, \nonumber
\end{eqnarray}
\noindent where:
\begin{eqnarray}
  \bF(x) = \left|\begin{array}{c}
                             \psi_1(x) \\
                             \psi_2(x)
                           \end{array}\right|\;,
\end{eqnarray}
\begin{eqnarray}
  \bB(x) = \frac{\hbar^2}{2m}\left[\begin{array}{cc}
                              0 & 1 \\
                              1 & 0
                            \end{array}\right]\;;
\end{eqnarray}
\begin{eqnarray}
  \bP(x)=\bY(x) = \frac{\hbar^2}{2m} q_y\left[\begin{array}{cc}
                              0 & 1 \\
                              -1 & 0
                            \end{array}\right]\;;
\end{eqnarray}
\begin{eqnarray}
  \bW(x) = \left[\begin{array}{cc}
            V_0-E & \frac{\hbar^2}{2m}q_y^2 \\
            \frac{\hbar^2}{2m}q_y^2 & V_0-E
          \end{array}\right].
\end{eqnarray}
The differential system (\ref{Eqmaestra}) defines the matrix differential operator ${\mathbf{L}}(x)$. Like the field $\bF(x)$, the linear differential form $\bA(x)=\left[\bB(x) \cdot d\bF(x)/dx + \bP(x) \cdot \bF(x)\right]$ associated to the operator ${\mathbf{L}}(x)$ is a useful magnitude for the transfer matrix method because it is continuous for every $x$ along the heterostructure \cite{36}.\\

The linearly independent(LI) solutions of the differential system (\ref{Eqmaestra}) can be expressed in the form of exponentials \cite{44,45}

\begin{eqnarray}
\label{Soluciones-LI-general-DCG} \bF (x)= \bF_{0} \, e^{ik\, x}\;,
\end{eqnarray}
\noindent where the eigenvalues $k$ and the corresponding amplitudes $\bF_{0}$ are the solutions of the Quadratic Eigenvalue Problem (QEP) \cite{37,46} that results when the LI solution $\bF (x)$ is substituted into the differential system. For our differential system (\ref{Eqmaestra}) we obtain the set of eigenvalues  $K=\{k_\ell, \ell=1,2,3,4\}$ given by:
\begin{eqnarray}
\label{K1}  k_{1} &=& +\sqrt{\frac{2m}{\hbar^2}(E-V_0)-q_y^2}, \\
\label{K2}  k_{2} &=& -k_{1}, \\
\label{K3}  k_{3} &=& - i\sqrt{\frac{2m}{\hbar^2}(E-V_0)+q_y^2}, \\
\label{K4}  k_{4} &=& - k_{3},
\end{eqnarray}
\noindent and the corresponding amplitudes:
\begin{eqnarray}\label{SL-barr-DCG}
 \bF_{\ell 0} =  \left|\begin{array}{c}
                    \frac{\hbar^2}{2m}(q_y+i k_{\ell })^2 \\
                    E-V_0
                \end{array}\right|;\;\;\ell=1,2,3,4\;.
\end{eqnarray}
The above expressions for eigenvalues $k_{\ell }$ and amplitudes $\bF_{\ell 0}$ were written for a barrier domain. For domains without electrostatic potentials they can be straightforwardly obtained by simply setting $V_0 = 0$.\\

For each eigenfunction $\bF_{\ell} (x)= \bF_{\ell0} \, e^{ik_{\ell}\, x}$ the corresponding linear differential form $\bA_{\ell}(x)$ can be expressed in the form $\bA_{\ell} (x)= \bA_{\ell0} \, e^{ik_{\ell}\, x}$ for both barrier and well domains, see appendix \ref{App-A}.\\

The solution of Equation (\ref{Eqmaestra}) for a single domain $\mu$ (internal or external) of our heterostructure can be expressed by means of a linear combination of the four eigenfunctions:

\begin{equation}\label{solg-R}
   \bF (\mu:x)=\sum_{\ell=1}^4 a_{\ell}(\mu) \bF_{\ell}(\mu:x),
\end{equation}

\noindent and the corresponding linear differential form:

\begin{equation}\label{solgA-R}
   \bA (\mu:x)=\sum_{\ell=1}^4 a_{\ell}(\mu) \bA_{\ell}(\mu:x),
\end{equation}

where $a_{\ell}$; $\ell=1,2,3,4$ is a set of expansion coefficients.\\

With the eigenfunctions available, the hybrid compliance-stiffness matrix or simply the hybrid matrix $\bH$ can be expressed as in  \cite{37}:

\begin{eqnarray}\label{H-zyzo}
  \bH(x;x_0) = \left[\begin{array}{cccc}
                               \mathbf{F}_{1}(x_0) & \mathbf{F}_{2}(x_0) & \mathbf{F}_{3}(x_0) & \mathbf{F}_{4}(x_0) \\
                               \mathbf{A}_{1}(x) &  \mathbf{A}_{2}(x) &  \mathbf{A}_{3}(x) &  \mathbf{A}_{4}(x)
                             \end{array}\right] \cdot \nonumber\\
                             \left[\begin{array}{cccc}
                               \mathbf{A}_{1}(x_0) & \mathbf{A}_{2}(x_0) & \mathbf{A}_{3}(x_0) & \mathbf{A}_{4}(x_0) \\
                               \mathbf{F}_{1}(x) &  \mathbf{F}_{2}(x) &  \mathbf{F}_{3}(x) &  \mathbf{F}_{4}(x)
                             \end{array}\right]^{-1}.
\end{eqnarray}
Here the zonal argument $\mu$ was suppressed to simplify the notation. After some algebraic manipulation of the right hand side of (\ref{H-zyzo}) we obtain the matrix $\bH$ in terms of the width $d=x-x_0$ of the domain $\mu$:

\begin{eqnarray}\label{H-d}
  \bH(d) = \left[\begin{array}{llll}
                               \bF_{1 0} & \bF_{2 0} & \bF_{3 0}\;e^{ik_{3}(-d)} & \bF_{4 0}\;e^{ik_{4}(-d)} \\
                               \bA_{10}\;e^{ik_{1}(d)} &  \bA_{20}\;e^{ik_{2}(d)} &  \bA_{30} &  \bA_{40}
                             \end{array}\right] \cdot \nonumber\\
                             \left[\begin{array}{cccc}
                               \bA_{10} & \bA_{20} & \bA_{30}\;e^{ik_{3}(-d)} & \bA_{40}\;e^{ik_{4}(-d)} \\
                                \bF_{1 0}\;e^{ik_{1}(d)} &  \bF_{2 0}\;e^{ik_{2}(d)} &  \bF_{3 0} &  \bF_{4 0}
                             \end{array}\right]^{-1}.
\end{eqnarray}

In order to calculate the transmittance in our GBG superlattice applying the hybrid matrix method, we need calculate the matrix $\bH$ of the whole superlattice. Using definition given in Reference \cite{37}, this matrix relates the magnitudes $\mathbf{F}(x)$ and $\mathbf{A}(x)$ at the ends $x_L$ and $x_R$ of the heterostructure:

\begin{eqnarray}
\label{matriz-H-Heterostructure}
  \begin{array}{|c|}
    \mathbf{F}(\mathrm{L}: x_L) \\
     \mathbf{A} (\mathrm{R}:x_R)
   \end{array}
   = \bH(x_R;x_L)\;\cdot \begin{array}{|c|}
                           \mathbf{A} (\mathrm{L}: x_L) \\
                           \mathbf{F}(\mathrm{R}:x_R)
                         \end{array}\;,
\end{eqnarray}
Matrix $\bH(x_R;x_L)$ can be calculated using the expression (\ref{H-d}) for the internal domains of the heterostructure (barriers and wells) and applying the composition rule given in appendix \ref{App-B} which define a product denoted here by the symbol $\odot$. For practical purposes we should determine the hybrid matrix $\bH_{uc}$ corresponding to a unit cell  of our superlattice, see Figure \ref{Fig2}:
\begin{equation}\label{Huc1}
    \bH_{uc}=\bH_b \odot \bH_w.
\end{equation}
Then for two barriers we have:
\begin{equation}\label{Huc2}
    \bH_{2b}=\bH_{uc} \odot \bH_b,
\end{equation}
\noindent and in general for $m$ barriers:
\begin{equation}\label{Huc3}
    \bH_{mb}=\bH_{uc} \odot \bH_{(m-1)b};\;\; m\geq2,
\end{equation}
\noindent where the subscripts $b$, $w$ in $\bH_b$, $\bH_w$ indicates that the hybrid matrix (\ref{H-d}) is calculated for a barrier, a well respectively. If the supperlattice contains $m$ barriers then $\bH(x_R;x_L)=\bH_{mb}$.\\

Finally, to calculate the transmittance we take into account the following two ideas:

\begin{enumerate}
  \item It is a common practice to choose a reduced basis at $x_L/x_R$, that is, the basis tending to unity at $x_L/x_R$. We selected the basis: $\bF_{\ell} (x)=\bF_{\ell0} \, e^{ik_{\ell}\, (x-x_L)}$ for left external domain $(\mathrm{L})$, and $\bF_{\ell} (x)=\bF_{\ell0} \, e^{ik_{\ell}\, (x-x_R)}$ for the right one $(\mathrm{R})$.\\

  \item Considering the same boundary conditions applied in the previous section, we set the value for some expansion coefficients given in (\ref{solg-R}-\ref{solgA-R}) for $\mu\equiv \{\mathrm{L}, \mathrm{R}\}$. That is $ a_{2}(\mathrm{R})=0$ since there is no reflected wave in the right external domain $(\mathrm{R})$ and $a_{3}(\mathrm{R})=0$, $a_{4}(\mathrm{L})=0$ to avoid nonphysical solutions as $x\rightarrow\pm\infty$.
\end{enumerate}

Then, applying the above boundary conditions and using the reduced basis in the expressions (\ref{solg-R}-\ref{solgA-R}); and substituting them in (\ref{matriz-H-Heterostructure}) we obtain, after some algebra, the equation:
\begin{eqnarray}
\fl
  \left|\begin{array}{c}
     a_2({\mathrm{L}})/a_1({\mathrm{L}}) \\
     a_3({\mathrm{L}})/a_1({\mathrm{L}}) \\
     a_1({\mathrm{R}})/a_1({\mathrm{L}}) \\
     a_4({\mathrm{R}})/a_1({\mathrm{L}})
  \end{array}\right| = \left[\mathbf{M}_1-\bH(x_R,x_L)\cdot \mathbf{M}_2\right]^{-1}\cdot\left\{\bH(x_R,x_L)\cdot \left|\begin{array}{c}
                                                                                                                                  \bA_{10} \\
                                                                                                                                  \bcero_{2\times1}
                                                                                                                                \end{array}\right|-\left|\begin{array}{c}
                                                                                                                                                           \bF_{10} \\
                                                                                                                                                           \bcero_{2\times1}
                                                                                                                                                         \end{array}\right|\right\},
\end{eqnarray}
\noindent where:
\begin{eqnarray}
  \mathbf{M}_1 = \left(\begin{array}{cccc}
                     \bF_{20} & \bF_{30} & \bcero_{2\times1} & \bcero_{2\times1} \\
                     \bcero_{2\times1} & \bcero_{2\times1} & \bA_{10} & \bA_{40}
                   \end{array}\right);
\end{eqnarray}
\begin{eqnarray}
  \mathbf{M}_2 = \left(\begin{array}{cccc}
                       \bA_{20} & \bA_{30} & \bcero_{2\times1} & \bcero_{2\times1} \\
                       \bcero_{2\times1} & \bcero_{2\times1} & \bF_{10} & \bF_{40} \\
                     \end{array}
                   \right).
\end{eqnarray}
In this case transmittance is calculated as: $T=\left|\frac{a_1({\mathrm{R}})}{a_1({\mathrm{L}})}\right|^2$.\\

Substituting the analytical expressions for $\bF_{\ell0}$, $\bA_{\ell0}$ and the eigenvalues $k_{\ell}$ in (\ref{H-d}) we obtained that the determinant of this matrix $\bH(d)$ equals unity regardless of the electron energy, the angle of incidence, the barrier height or the domain width. Then, to verify the numerical stability of the hybrid matrix method, the determinant of the matrix $\bH(x_R;x_L)$ (corresponding to the whole superlattice) was calculated for different values of this parameters, as well as for different number of barriers.

\section{Results}

\begin{figure}
\centering
\includegraphics[scale=0.3]{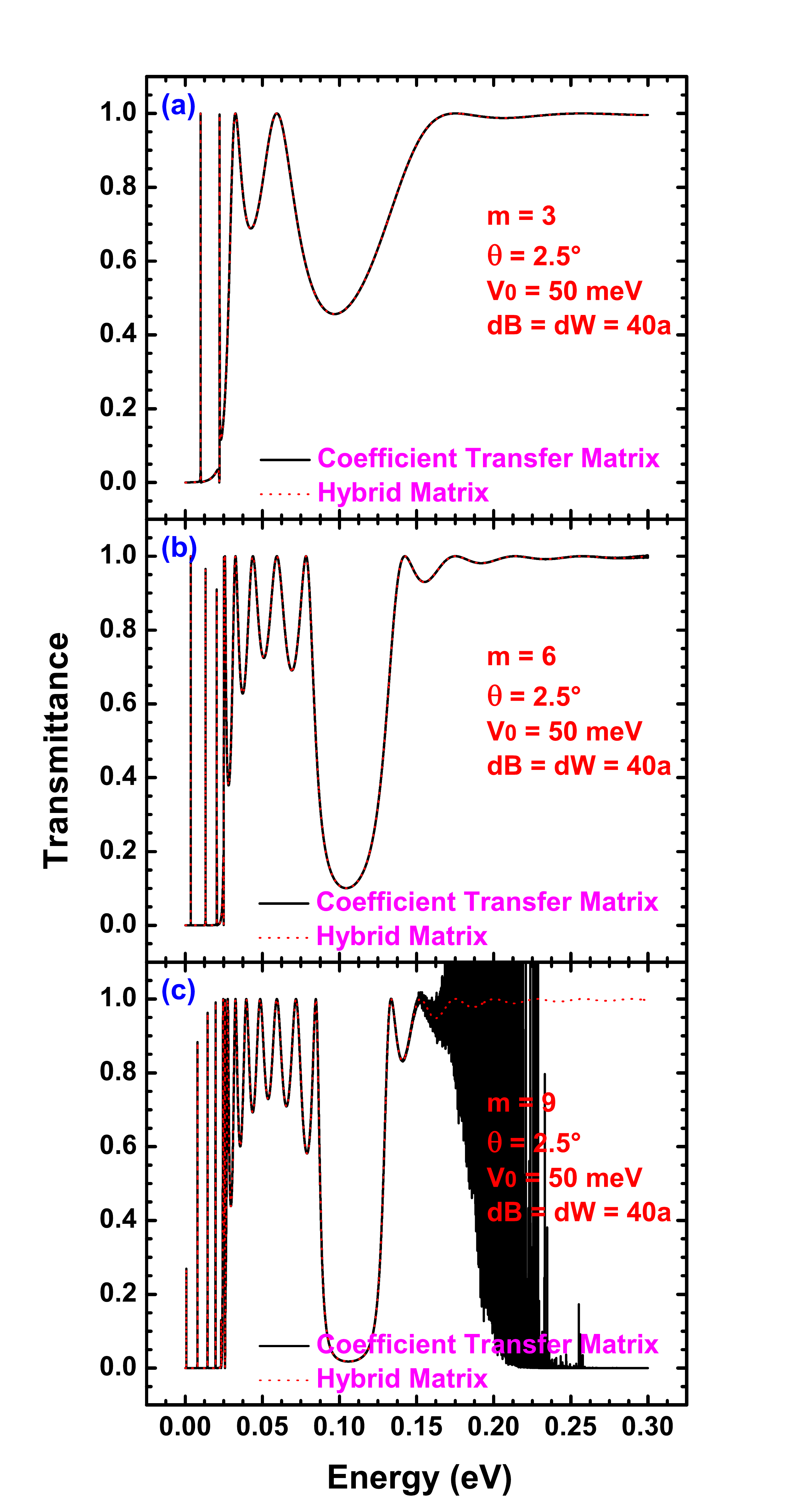}
\caption{\label{Fig3} Comparison of the transmittance as function of energy for the coefficient transfer matrix (solid black line) and the hybrid matrix (dashed-red line) for a GBG superlattice. The width of the barriers and wells is $40a$, the angle of incidence is $2.5^{\circ}$, the height of barriers is $50$ meV. Figs. (a), (b) and (c) correspond to number of barriers 3, 6 and 9, respectively.}
\end{figure}

Our main purpose is to show that the practical application of the common coefficient transfer matrix $(\bK)$ to the study of transport properties (e.g., transmission coefficients) in GBG superlattices is hampered by the $\Omega d$ problem, and this numerical degradation can be avoided applying the proposed hybrid matrix method which gives stable and reliable results for the same structure. To this end, we calculated transmittance spectra using both methods and used the matrix determinant as a test of the numerical accuracy in the calculations. The corresponding transmittance spectra were compared considering the growth in the energy of the incident electrons, the angle of incidence, the barrier height and thickness and number of barriers. Fig. \ref{Fig3} presents the transmittance as a function of the energy for a relatively low angle of incidence $(\theta=2.5^{\circ})$ at which the Fano resonances can manifest. In this case we considered a GBG superlattice with barrier and well widths of $dB=dW=40a$ and barrier height $V_0=50$ meV. Here $a$ is the carbon-carbon distance in graphene, which is equal to 0.142 nm. In addition, a different number $m$ of barriers was considered, see Figs. \ref{Fig3} (a), (b) and (c) for $m=3$, $m=6$ and $m=9$ respectively. As can be seen, both methods give similar results for three and six barriers but for nine barriers structure it appears a remarkable difference above 150 meV indicating the numerical degradation of calculation involving the $\bK$ matrix method. The signs of the $\Omega d$ problem in the system of nine barriers is due in part to the cumulative nature of this numerical degradation and partly to the energy increase. The latter causes the exponential argument $\beta_x\;d$ grows and sooner or later, as this product grows, the difference in order of magnitude between the very large and the very small exponentials in (\ref{matrix-P}) exceeds the accuracy with which our computer is working and numerical degradation sets. On the other hand, from expression (\ref{Matrix-K}) it is clear that the number of matrices involved in this repeated matrix multiplication is proportionally to the number of constituent domains (barriers and well). This repeated matrix multiplication means a sequence of calculations prone to roundoff errors which may accumulate \cite{37}. When this happen the numerical degradation can dominates the calculations thus giving a very inaccurate final result. This is the case of transmittance in the nine barriers structure, see Fig. \ref{Fig3} (c). The transmittance spectra calculated using the hybrid matrix method show a well behaved in the whole range of energy even for the nine barriers system, indicating the numerical stability of this matrix method.\\

\begin{figure}
\centering
\includegraphics[scale=0.3]{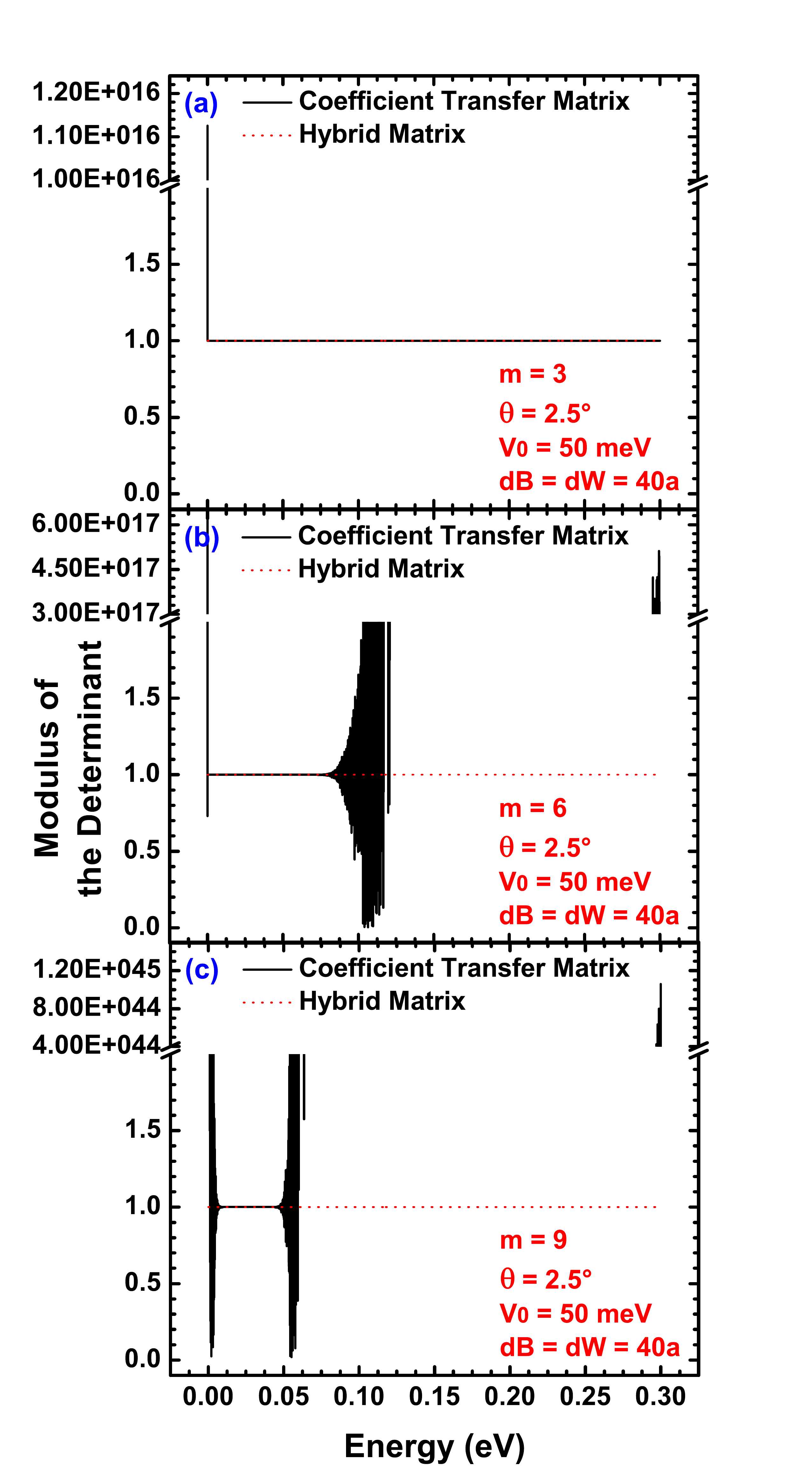}
\caption{\label{Fig4} Modulus of the determinant versus the energy for a GBG superlattice, the parameters are the same as in Fig. \ref{Fig3}}
\end{figure}

\begin{figure}
\centering
\includegraphics[scale=0.6]{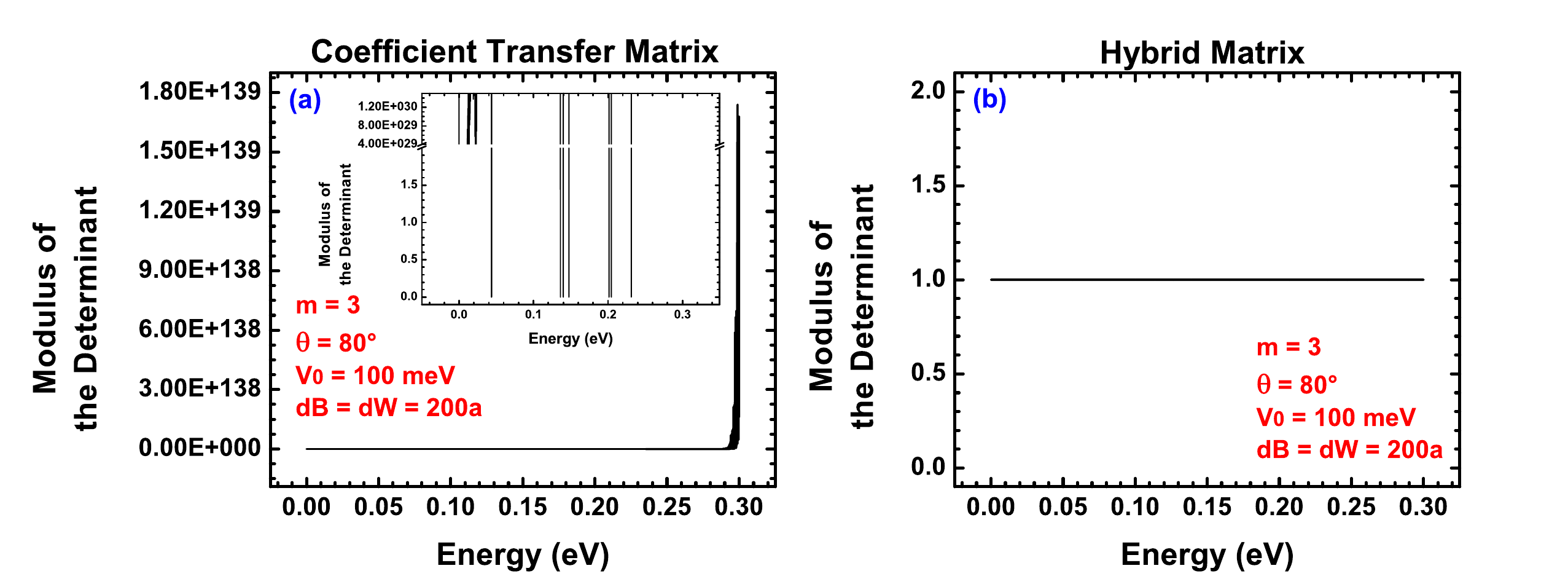}
\caption{\label{Fig5} Modulus of the determinant versus the energy for a GBG superlattice. The width of the barriers and wells is $200a$, the angle of incidence is $80^{\circ}$, the height of barriers is $100$ meV and the number of barriers is 3.}
\end{figure}

Figure 4 show the modulus of the determinant of the whole structure matrix for the GBG structures analyzed in Fig. \ref{Fig3}. In contrast to the hybrid matrix whose determinant is invariably unimodular, the determinant of the coefficient transfer matrix takes values quite different from exactly one, being in some cases several orders of magnitude bigger than 1.0. Fig. \ref{Fig4} (b) and (c) indicates that this behavior is more pronounced when the number of barriers increases. It should be noted that the deviation experimented by the determinant of matrix $\bK$ near the zero energy is the result of the eigenfunction basis (Eqs. \ref{Eigenfunction-q} and \ref{Eigenfunction-beta}) used to construct this matrix: as the energy $E$ appear in the denominator of the expressions defining the parameters $v_{\pm}^q$ and $v_{\pm}^\beta$ they can be very large for energies near to zero in the regions without electrostatic potentials. This deviation can be avoided changing the eigenfunction basis. Fig \ref{Fig5}(b) indicates that the determinant of matrix $\bH$ can be unimodular even at a high angle of incidence and significantly higher values of the parameters $dB$, $dW$ and $V_0$. Here it is important to mention that in principle we can expect a correspondence between the energy range of numerical instabilities in the transmission probability and the energy range at which the determinant deviates from the unity. However, this is not the case because in the transmission probability only four elements of the coefficient transfer matrix enter in the computation, while in the determinant all elements of the matrix are involved, see Figs. \ref{Fig3}b and \ref{Fig4}b. So, some elements that can contribute to the deviation of the determinant are not necessarily involved in the determination of the transmittance, and hence the correspondence between these quantities is not fulfilled at all.   

\begin{figure}
\centering
\includegraphics[scale=0.4]{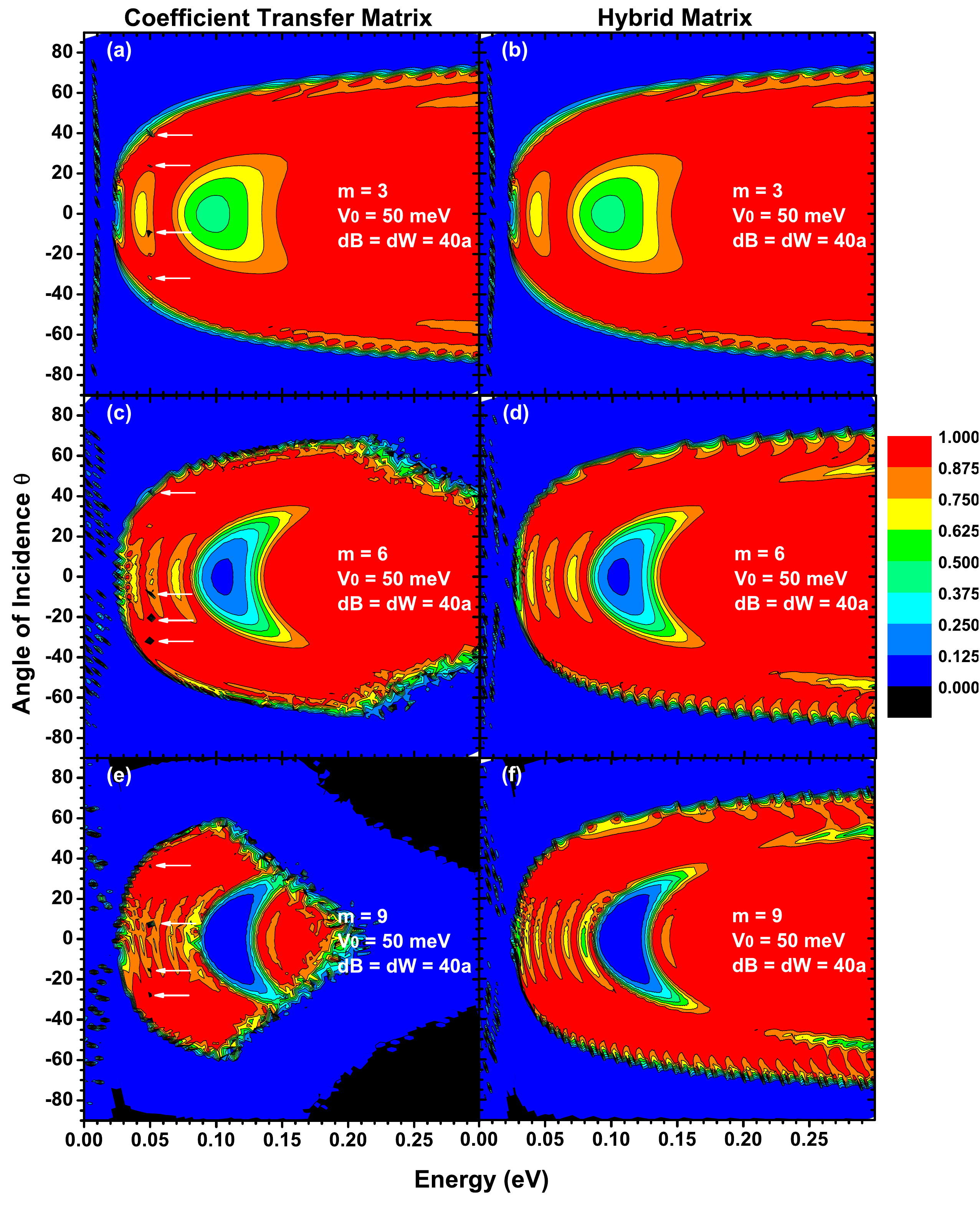}
\caption{\label{Fig6} (Color Online) Contour plots of the transmittance as function of the energy and the angle of incidence for a GBG superlattice. First column corresponds to the coefficient transfer matrix, and second column to the hybrid matrix. The first ((a) and (b)), second ((c) and (d)), and third ((e) and (f)) rows correspond to 3, 6 and 9 barriers. We are using the same parameters as in Fig. \ref{Fig3}.}
\end{figure}

\begin{figure}
\centering
\includegraphics[scale=0.4]{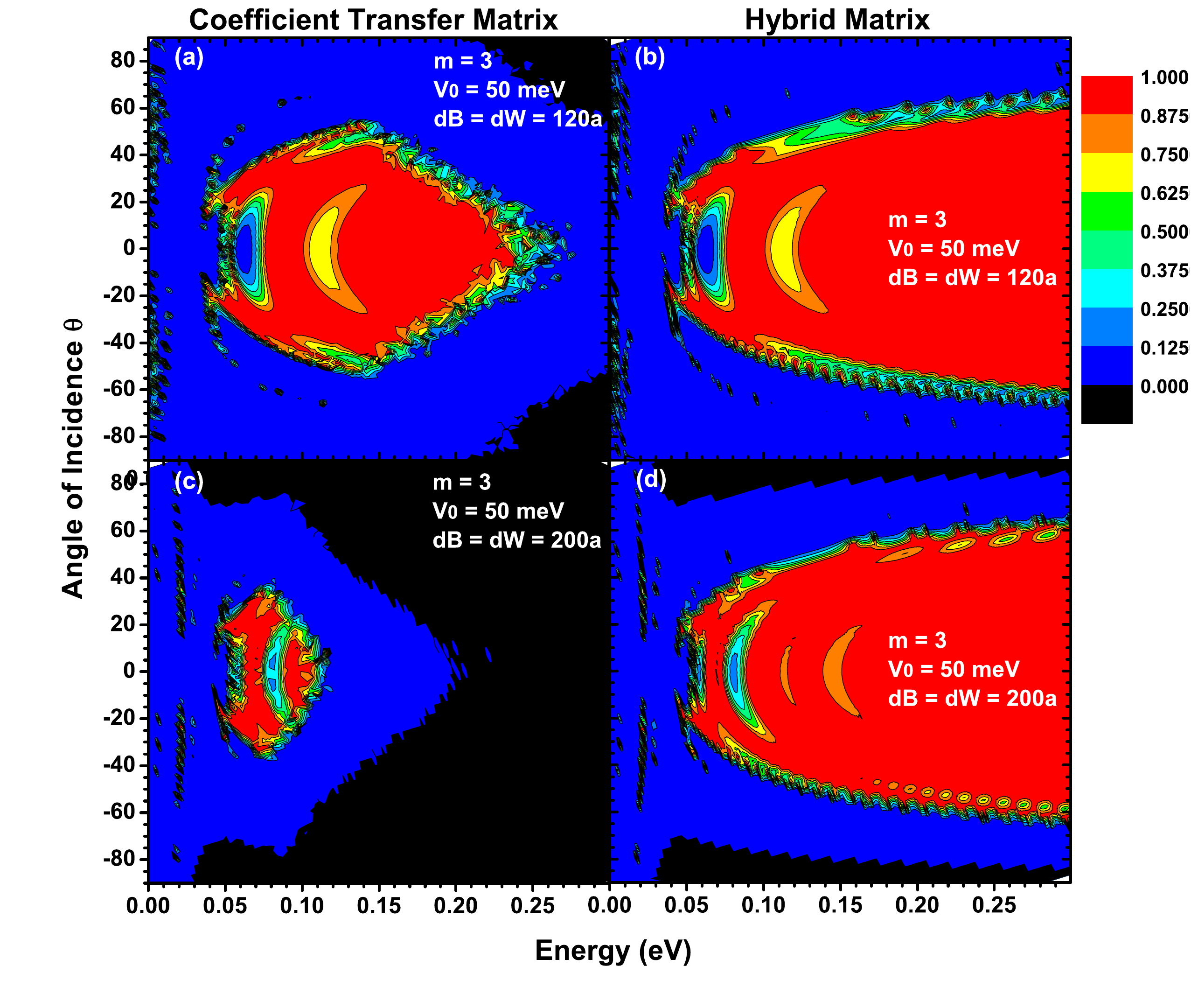}
\caption{\label{Fig7} (Color Online) Contour plots of the transmittance as function of the energy and the angle of incidence for a GBG superlattice. First column corresponds to the coefficient transfer matrix, and second column to the hybrid matrix. The number of barriers is $m=3$ and the height of the barriers is $V_{0}=50$ meV. The widths of wells and barriers are: $120a$ ((a) and (b)), and $200a$ ((c) and (d)).}
\end{figure}

A better perspective of the transmission properties is obtained from the transmittance contour plots given in Figure \ref{Fig6}. The color scale indicates Transmittance values as function of the energy and the angle of incident electrons. The Transmittance was calculated using both matrix method ($\bK$  and $\bH$) for $dB=dW=40a$, $V_0=50$ meV and the number of barriers $m=3$, $m=6$ and $m=9$, as in Figure \ref{Fig3}. Black dot market by white arrows in Fig. \ref{Fig6}a correspond to the deviation near the zero energy which was explained before. Figure \ref{Fig6}c shows that degradation affecting the calculations performed with the matrix $\bK$ is more likely when the energy and the angle of the incident electron increase, and Figure \ref{Fig6}e shows that this behavior is more pronounced when the number of barriers increases. The dramatic disappearance of the contour plot in Figure \ref{Fig6}e caused by the $\Omega d$ problem, is also more notable when the width of the barriers and wells increases as is shown in Figure \ref{Fig7}. In these graphs the numerical degradation causes loss of physical information concerning the Fano resonances in regions of perfect transmission as well as information about the anti-Klein tunneling. Contrary to this, the calculations performed with the matrix $\bH$ are stable and reliable for all the electron energies and angles of incidence covered as well as for all the domain widths and the number of barriers considered. Figures \ref{Fig6} and \ref{Fig7} indicates that the numerical degradation affecting these practical applications of the coefficient transfer matrix method is quite sensitive to the energy of the electrons, the angle of incidence and the size of the system.\\

\section{Conclusions}

In summary, we studied the propagation of Dirac electrons in GBG superlattices using both the coefficient transfer matrix method and the hybrid matrix method. We have shown that the practical application of the coefficient transfer matrix $(\bK)$ to study the transport properties (e.g., transmission coefficients) in GBG superlattices is hampered by a numerical degradation (the so called $\Omega$d problem), which is more likely when the energy of the electrons, the angle of incidence and the size of the system grow. This numerical degradation causes loss of physical information concerning the Fano resonances in the regions of perfect transmission as well as information about the anti-Klein tunneling.\\

To avoid the $\Omega$d problem we presented a straightforward procedure based in the hybrid matrix method ($\bH$). This matrix method has being developed from the matrix Sturm-Liouville version of the effective Hamiltonian describing the motion of charge carriers in bilayer graphene. The accuracy and the numerical stability of the hybrid method was showed for all the electron energies and angles of incidence covered as well as for all the domain widths and the number of barriers considered. This important result allows to track the Fano resonances for all relevant parameters of the GBG superlattice and also opens a door to discern the contribution of the Fano resonances in the transport properties, and in this way an excellent opportunity to test this exotic phenomenon from the experimental standpoint. \\

We also showed that the matrix determinant can be used as an effective test of the numerical accuracy in real calculations.

\section{Acknowledgments}

J. A. B.-T. acknowledges to CONACYT-M\'{e}xico for the financial support for the doctoral studies. 

\section{Appendix}

\appendix

\subsection{\label{App-A} Calculus of the linear differential form}

Applying the definition for $\bA(x)$, we have:

\begin{equation}\label{}
    \bA_{\ell}(x)=\bB(x) \cdot \frac{d\bF_{\ell}(x)}{dx} + \bP(x) \cdot \bF_{\ell}(x).
\end{equation}
Then, substituting the expression for $\bB(x)$, $\bP(x)$ and the eigenfunctions $\bF_{\ell} (x)= \bF_{\ell0} \, e^{ik_{\ell}\, x}$, with:

\begin{eqnarray}\label{}
 \bF_{\ell 0} =  \left|\begin{array}{c}
                    \frac{\hbar^2}{2m}(q_y+i k_{\ell })^2 \\
                    E-V_0
                \end{array}\right|;\;\;\ell=1,2,3,4\;,
\end{eqnarray}
\noindent we obtained, after a simple algebra:
\begin{eqnarray}\label{Al-1}
  \bA_{\ell}(x) = \left|\begin{array}{c}
                         \frac{\hbar^2}{2m} (E-V_0)(q_y+i k_{\ell })\\
                         \\
                         -\frac{\hbar^2}{2m} (E-V_0)(q_y+i k_{\ell })
                         \end{array}\right|\;e^{ik_{\ell }x};\;\;\ell=1,2;
\end{eqnarray}
\begin{eqnarray}\label{Al-2}
  \bA_{\ell }(x) = \left|\begin{array}{c}
                            \frac{\hbar^2}{2m} (E-V_0)(q_y+i k_{\ell })\\
                            \\
                            \frac{\hbar^2}{2m} (E-V_0)(q_y+i k_{\ell })
                            \end{array}\right|\;e^{ik_{\ell }x};\;\;\ell=3,4\;.
\end{eqnarray}
For domains without electrostatic potentials the above expressions must modified by simply setting $V_0 = 0$. As can be seen expressions (\ref{Al-1}-\ref{Al-2}) can be cast in a concise form as:
\begin{equation}\label{}
    \bA_{\ell} (x)=\bA_{\ell0} \, e^{ik_{\ell}\, x}.
\end{equation}

\subsection{\label{App-B} Composition rule for the hybrid matrix}

\begin{figure}
\centering
\includegraphics[width=0.3\textwidth]{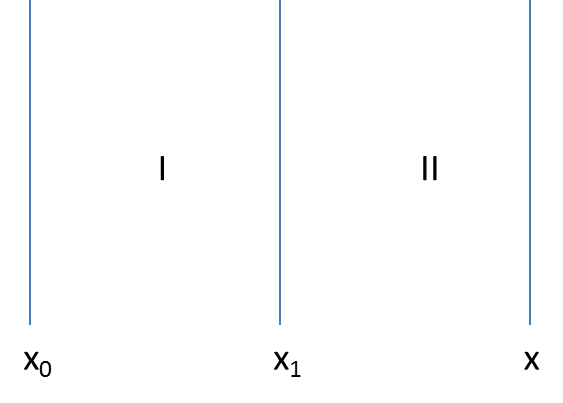}\\
\caption{\label{Fig8} A general scheme of an heterostructure composed by two adjacent domains, showing the locations of its interfaces.}
\end{figure}

Suppose we know the $4\times4 $ hybrid matrix $\bH^I (x_1,x_0)$ for the domain I and the corresponding one for the domain II $(\bH^{II} (x,x_1))$, see Figure \ref{Fig8}. Then the hybrid matrix $\bH^{III} (x,x_0)$ for the whole heterostructure can be obtained by means of the composition rule of the hybrid matrix \cite{37}. Taking into account the continuity of the field $\bF(x)$ and the associated linear form $\bA(x)$ in $x_1$ we have:

\begin{eqnarray}\label{regla-composicion-H}
  \bH^{III}_{11} &=& \bH^I_{11}+\bH^I_{12}\cdot\bH^{II}_{11}\cdot \left[\bI_2-\bH^I_{22}\cdot\bH^{II}_{11}\right]^{-1}\cdot\bH^I_{21}\; \nonumber\\
  \bH^{III}_{12} &=& \bH^I_{12}\cdot\left[\bI_2+\bH^{II}_{11}\cdot \left[\bI_2-\bH^I_{22}\cdot\bH^{II}_{11}\right]^{-1}\cdot\bH^I_{22}\right]\cdot\bH^{II}_{12};\nonumber\\
  \bH^{III}_{21} &=& \bH^{II}_{21}\cdot\left[\bI_2-\bH^I_{22}\cdot\bH^{II}_{11}\right]^{-1}\cdot\bH^I_{21};\nonumber \\
  \bH^{III}_{22} &=& \bH^{II}_{22}+\bH^{II}_{21}\cdot\left[\bI_2-\bH^I_{22}\cdot\bH^{II}_{11}\right]^{-1}\cdot\bH^I_{22}\cdot\bH^{II}_{12}.
\end{eqnarray}
The above expressions involve the four $2\times2$ blocks of the matrices $\bH^I (x_1,x_0)$, $\bH^{II} (x,x_1)$ and $\bH^{III} (x,x_0)$. We denoted by $\bI_2$ the unit matrix or order 2. We can now define the product, denoted $\odot$, namely:
\begin{equation}\label{}
    \bH^{III}(x,x_0)=\bH^{II}(x,x_1)\odot\bH^I(x_1,x_0),
\end{equation}
\noindent by de composition rule (\ref{regla-composicion-H}).

\pagebreak

\section*{References}
\begin{thebibliography}{99}
\bibitem{1} K. S. Novoselov, A. K. Geim, S. V. Morozov, D. Jiang, Y. Zhang, S. V. Dubonos, I. V. Grigorieva and A. A. Firsov, \textit{Science} \textbf{306} (2004) 666.
\bibitem{2} K. S. Novoselov, A. K. Geim, S. V. Morozov, D. Jiang, M. I. Katsnelson, I. V. Grigorieva, S. V. Dubonos and A. A. Firsov, \textit{Nature} \textbf{438} (2005) 197.
\bibitem{3} K. S. Novoselov, Z. Jiang, Y. Zhang, S. V. Morozov, H. L. Stormer, U. Zeitler, J. C. Maan, G. S. Boebinger, P. Kim and A. K. Geim, \textit{Science} \textbf{315} (2007) 1379.
\bibitem{4} Y. Zhang, J.-W. Tan, H. L. Stormer and P. Kim, \textit{Nature}, \textbf{438} (2005) 201.
\bibitem{5} M. I. Katsnelson, \textit{Eur. J. Phys. B} \textbf{51} (2006) 157.
\bibitem{6} M. I. Katsnelson, K. S. Novoselov and A. K. Geim, \textit{Nat. Phys.} \textbf{2} (2006) 620.
\bibitem{7} A. F. Young, and P. Kim, \textit{Nat. Phys.} \textbf{5} (2009) 222.
\bibitem{8} C. Bai, and X. Zhang, \textit{Phys. Rev. B} \textbf{76} (2007) 075430.
\bibitem{9} C.-H. Park, L. Yang, Y.-W. Son, M. L. Cohen and S. G. Louie, \textit{Nat. Phys.} \textbf{4} (2008) 217.
\bibitem{10} C.-H. Park, Y.-W. Son, L. Yang, M. L. Cohen and S. G. Louie, \textit{Phys. Rev. Lett.} \textbf{103} (2009) 046808.
\bibitem{11} M. Barbier, P. Vasilopoulos, and F. M. Peeters, \textit{Phys. Rev. B} \textbf{81} (2010) 075438.
\bibitem{12} Y.-X. Li, \textit{J. Phys.: Condens. Matter} \textbf{22} (2010) 015302.
\bibitem{13} Q.-S. Wu, S.-N. Zhang, and S.-J. Yang, \textit{J. Phys.: Condens. Matter} \textbf{20} (2008) 485210.
\bibitem{14} A. Isacsson, L. M. Jonsson, J. M. Kinaret, and M. Jonson, \textit{Phys. Rev. B} \textbf{77} (2008) 035423.
\bibitem{15} J. M. Pereira, Jr., V. Mlinar, F. M. Peeters, and P. Vasilopoulos, \textit{Phys. Rev. B} \textbf{74} (2006) 045424.
\bibitem{16} I. Rodr\'{i}guez-Vargas, J. Madrigal-Melchor, and O. Oubram, \textit{J. Appl. Phys.} \textbf{112} (2012) 073711.
\bibitem{17} J. A. Briones-Torres, J. Madrigal-Melchor, J. C. Mart\'{i}nez-Orozco and I. Rodr\'{i}guez-Vargas, \textit{Superlattice. Microst.} \textbf{73} (2014) 98.
\bibitem{18} K. S. Novoselov, E. McCann, S. V. Morozov, V. I. Fal'ko, M. I. Katsnelson, U. Zeitler, D. Jiang, F. Schedin and A. K. Geim, \textit{Nature} \textbf{2} (2006) 177.
\bibitem{19} M. S. Dresselhaus and G. Dresselhaus, \textit{Adv. Phys.} \textbf{51} (2002) 1.
\bibitem{20} J. C. Charlier, J. P. Michenaud, X. Gonze, \textit{Phys. Rev. B} \textbf{46} (1992) 4531.
\bibitem{21} A. K. Geim and K. S. Novoselov, \textit{Nat. Mater.} \textbf{6} (2007) 183.
\bibitem{22} E. McCann and V. I. Fal'ko, \textit{Phys. Rev. Lett.} \textbf{96} (2006) 086805.
\bibitem{23} E. McCann, D. S. L. Abergel and V. I. Fal'ko, \textit{Solid State Commun.} \textbf{143} (2007) 110.
\bibitem{24} E. McCann and M. Koshino, \textit{Rep. Prog. Phys.} \textbf{76} (2013) 056503.
\bibitem{25} T. Ohta, A. Bostwick1, T. Seyller, K. Horn and E. Rotenberg, \textit{Science} \textbf{313} (2006) 951.
\bibitem{26} E. V. Castro, K. S. Novoselov, S. V. Morozov, N. M. R. Peres, J. M. B. Lopes dos Santos, J. Nilsson, F. Guinea, A. K. Geim, and A. H. Castro Neto, \textit{Phys. Rev. Lett.} \textbf{99} (2007) 216802.
\bibitem{27} Y. Zhang, T. -T. Tang, C. Girit, Z. Hao, M. C. Martin, A. Zettl, M. F. Crommie, Y. R. Shen and F. Wang, \textit{Nature} \textbf{459} (2009) 820.
\bibitem{28} S. Mukhopadhyay, R. Biswas and C. Sinha, \textit{J. Appl. Phys.} \textbf{110} (2011) 014306.
\bibitem{29} S. Mukhopadhyay, R. Biswas and C. Sinha, \textit{Phys. Lett. A} \textbf{375} (2011) 2921.
\bibitem{30} C. Sinha and R. Biswas, \textit{Phys. Rev. B} \textbf{84} (2011) 155439.
\bibitem{31} A. E. Miroshnichenko, S. Flach and Y. S. Kivshar, \textit{Rev. Mod. Phys.} \textbf{82} (2010) 2257.
\bibitem{32} B. Luk'yanchuk, N. I. Zheludev, S. A. Maier, N. J. Halas, P. Nordlander, H. Giessen, and C. T. Chong, \textit{Nat. Mater.} \textbf{9} (2010) 707.
\bibitem{33} Y. Wang, \textit{J. Appl. Phys.} \textbf{116} (2014) 164317.
\bibitem{34} P. Yeh, \textit{Optical waves in layered media} (John Wiley y Sons, Inc., 2005).
\bibitem{35} P. Markos and C. M. Soukoulis, \textit{Wave Propagation: From Electrons to Photonic Crystals and Left-Handed Materials} (Princeton University Press, 2008).
\bibitem{36} R. P\'{e}rez-\'{A}lvarez and F. Garc\'{i}a-Moliner, \textit{Transfer Matrix, Green Function and related techniques: Tools for the study of multilayer heterostructures} (Universitat Jaume I, Castell\'{o}n de la Plana, Spain, 2004).
\bibitem{37} R. P\'{e}rez-\'{A}lvarez, R. Pernas-Salom\'{o}n and V. R. Velasco, \textit{SIAM J. Appl. Math.} \textbf{75} (2015) 1403.
\bibitem{38} E. L. Tan, \textit{IEEE T. Ultrason. Ferr.} \textbf{42} (1995) 525.
\bibitem{39} E. L. Tan, \textit{J. Acoust. Soc. Am.} \textbf{119} (2006) 45.
\bibitem{40} R. Pernas-Salom\'{o}n and R. P\'{e}rez-\'{A}lvarez, \textit{Prog. Electromagn. Res. M} \textbf{40} (2014) 79.
\bibitem{41} E. Maci\'{a}, \textit{Rep. Prog. Phys.} \textbf{69} (2006) 397.
\bibitem{42} R. Rodr\'{i}guez-Gonz\'{a}lez and I. Rodr\'{i}guez-Vargas, \textit{Physica E} \textbf{69} (2015) 177.
\bibitem{43} S. Mukhopadhyay, R. Biswas and C. Sinha, \textit{Phys. Status Solidi B} \textbf{247} (2010) 342.
\bibitem{44} V. Hurewicz, \textit{Lectures on ordinary differential equations} (The MIT Press, Cambridge, Massachusetts, 1958).
\bibitem{45} Y. Bibikov, \textit{General course on ordinary differential equations (in Russian)} (Leningrad University Press, 1981).
\bibitem{46} F. Tisseur and K. Meerbergen, \textit{SIAM Review} \textbf{43} (2001) 235.
\end{thebibliography}
\end{document}